\documentclass[a4paper,11pt]{article}
\pdfoutput=1 % if your are submitting a pdflatex (i.e. if you have
\usepackage{jcappub} % for details on the use of the package, please
\usepackage[T1]{fontenc} % if needed
\usepackage{xcolor}
\usepackage{graphicx}
\usepackage{mwe}
\usepackage{physics}
\usepackage{caption}
\usepackage{subcaption}
\usepackage{mathrsfs}
\usepackage{bm}
\usepackage[normalem]{ulem}

\newcommand{\obs}[1]{\hat{#1}}
\newcommand{\theory}[1]{#1^{\textrm{\scriptsize{theory}}}}
\newcommand{\sys}[1]{#1^{\textrm{\scriptsize{sys}}}}
\newcommand{\thetasysi}{\theta_{\textrm{\scriptsize{sys}},i}}
\newcommand{\kmax}{k_{\scriptsize{\textrm{max}}}}

\def\blapl{C_{\nabla^2\delta}}

\title{Neutrino mass constraints beyond linear order: cosmology dependence and systematic biases}

\author[a,b,1]{Aoife Boyle\note{Corresponding author.}}
\author[a]{and Fabian Schmidt}

\affiliation[a]{Max-Planck-Institut f\"{u}r Astrophysik, Karl-Schwarzschild-Str. 1, 85748 Garching bei M\"{u}nchen, Germany}
\affiliation[b]{CNRS \& Sorbonne Universit\'e, UMR 7095, Institut d'Astrophysique de Paris, 75014, Paris, France}

\emailAdd{boyle@iap.fr}
\emailAdd{fabians@mpa-garching.mpg.de}

\abstract{
  \noindent We demonstrate the impact on forecasted neutrino mass constraints of improving galaxy clustering and CMB lensing predictions from linear to next-to-leading-order power spectra. The redshift-space 1-loop power spectrum model we adopt requires an additional four free bias parameters, a velocity bias parameter and two new stochastic parameters. These additional nuisance parameters appreciably weaken the constraints on $M_\nu$. CMB lensing plays a significant role in helping to alleviate these degeneracies and tighten the final constraints. The constraint on the optical depth to reionisation $\tau$ has a strong effect on the constraint on $M_\nu$, but only when CMB lensing is included in the analysis to keep the degeneracies with the nuisance parameters under control. We also extract constraints when 1) using the BAO signature only as a distance probe, and 2) isolating the scale-dependence of the power spectrum, which, as shown in previous work, provides a cosmology-independent probe of $M_\nu$. All constraints except the latter remain strongly sensitive to the assumption of a flat $\Lambda$CDM universe. We perform an analysis of the magnitude of the shift introduced in the inferred $M_\nu$ value when neglecting nonlinear corrections, and show that, for a Euclid-like survey, this shift becomes roughly equal to the $1\sigma$ constraint itself even with a conservative cut-off scale of $k_\textrm{max}=0.1~h~\textrm{Mpc}^{-1}$. We also perform a calculation of the approximate expected bias in neutrino mass caused by not including the next, 2-loop order and expect a shift of only about 20\% of the $1\sigma$ error for $k_\textrm{max}=0.2~h~\textrm{Mpc}^{-1}$ in a Euclid-like survey. 
}

\begin{document}
\maketitle
\flushbottom 

\section{Introduction}

The promise of improved constraints on the sum of neutrino masses $M_\nu$ is one driving force behind planned large-scale structure surveys \cite[e.g.][]{amendola_cosmology_2016, laureijs_euclid_2011, green_wide-field_2012, tamura_prime_2016, desi_collaboration_desi_2016} and CMB experiments \cite[e.g.][]{the_simons_observatory_collaboration_simons_2019, abazajian_cmb-s4_2016}. At present, solar and atmospheric oscillation experiments impose a minimum $M_\nu$ (for the normal hierarchy, which is currently favoured at 3$\sigma$ \cite{globalfit, de_salas_2020_2020}) of just under 0.06 eV. The upper limits provided by cosmology still significantly outstrip all experimental alternatives, with the 2018 Planck data release asserting a 2$\sigma$ constraint\footnote{Note that the Planck constraints are quoted as a 2$\sigma$ upper limit, but the forecasted constraints we present in this paper are 1$\sigma$ uncertainties.} of $M_\nu<0.12$ eV by combining CMB and baryon acoustic oscillation (BAO) data \cite{planck_collaboration_planck_2019} for a flat $\Lambda$CDM+$M_\nu$ cosmology. However, the constraints on the neutrino mass from galaxy clustering and lensing measurements are inherently complex, and are limited by a number of factors including cosmological degeneracies \cite{allison_towards_2015, archidiacono_physical_2017, boyle_deconstructing_2018, boyle_understanding_2019, choudhury_updated_2020}, baryonic effects \cite{chung_baryonic_2020} and theoretical uncertainties related to the small-scale power spectrum \cite{baldauf_lss_2016, brinckmann_promising_2019, chudaykin_measuring_2019}.

In previous work \cite{boyle_deconstructing_2018, boyle_understanding_2019}, we have attempted to deconstruct forecasted neutrino mass constraints from future galaxy clustering and CMB surveys. Our aim has been to assess the robustness of expected constraints, considering in particular how they could be weakened by possible alterations to flat $\Lambda$CDM. We have demonstrated that constraints on the total neutrino mass, $M_\nu$, are heavily affected by possible small deviations in the assumed cosmology, even with powerful next-generation surveys. However, we have also shown that isolating the effects of neutrino free-streaming in the power spectrum as a constraining tool can provide cosmology-independent constraints. 

In the previous work cited above, we based our forecasts entirely in the linear regime. We assumed the Kaiser formula to describe the tree-level redshift-space galaxy power spectrum (see Section \ref{sec_methodology} for details). This formulation is accurate for quite a limited range of potentially usable scales, and takes a rather crude approach to galaxy bias and redshift-space distortions, and neglects any selection effects. We derived our constraint forecasts for $M_\nu$ assuming a minimum usable comoving scale $k_\textrm{max}=0.2~h~\textrm{Mpc}^{-1}$, but this is likely to be optimistic. Moving beyond the assumption of linearity provides a natural and important extension to our work. This is particularly important because a substantial contribution to the constraint on $M_\nu$ is provided by the suppression in the power spectrum on small scales, where non-linear corrections become increasingly significant. 

Recently, Ref.~\cite{desjacques_galaxy_2019} presented the first complete expression for the perturbative next-to-leading-order (NLO) or 1-loop power spectrum in redshift space including all selection effects
(see \cite{Scoccimarro:2004tg,Seljak:2011tx,Perko:2016puo,fonseca/etal:2018} for related derivations). Using the approach developed in \cite{mirbabayi_biased_2014}, the formalism presented orders the contributing terms in a consistent way and keeps the contributions appropriately general. It also includes the leading higher-derivative bias and velocity bias terms. This rigorous approach leads to a complex but complete final expression for the redshift-space power spectrum that makes minimal assumptions. We implement this formalism here. Ref.~\cite{desjacques_galaxy_2019} emphasises the importance of accounting for line-of-sight-dependent selection effects in the calculation of the galaxy power spectrum unless physical arguments can be made for neglecting them. We neglect selection effects for most of this article, but justify this and provide some extended calculations in Section \ref{subsec_selection_effects}. Another recent work, Ref.~\cite{agarwal_information_2020}, presented a Fisher analysis of the cosmological information available in the 1-loop power spectrum and tree-level galaxy bispectrum, after fully accounting for selection effects, for future galaxy surveys like Euclid \cite{amendola_cosmology_2016, laureijs_euclid_2011}. Here we present a similar analysis, but based on the power spectrum only and focused on the neutrino mass, and extending to consider a range of possible cosmologies. We discuss our results in the context of recent related work, such as \cite{chudaykin_measuring_2019}, in Section \ref{subsec_previous_work}.

We proceed as follows. In Section \ref{sec_methodology}, we discuss the implementation of the NLO galaxy and CMB lensing power spectra in our Fisher matrix code, our fiducial cosmology and surveys and our calculation methods. In Section \ref{sec_results} we provide results from the linear and NLO power spectra for three scenarios: using the full broadband power spectrum (we call this the `combined' calculation),
%\fs{in the galaxy clustering community, people typically use ``full shape'' for this. Shall we adopt that too instead of ``combined''?} \comment{The reasoning for using ``combined'' originally was to clearly separate the full power spectrum and the scale-dependent neutrino effect (both of which could kind of be called ``shape'', so I changed the latter to be ``free-streaming'' for the same reason). It also follows on from this deconstruction approach, where we sort of broke everything down and then put it back together in the first paper. I'd like to keep it for consistency but if you think ``full shape'' is much clearer we can change it.}, \fs{it's fine to keep consistency, also since ``full shape'' is not uniquely defined.}
using the BAO feature only, and using only the scale-dependent suppression of the power spectrum unique to neutrinos (which we call the `free-streaming' calculation). We also examine how CMB lensing can improve these constraints. We present all of our results in the form of forecasts for Euclid \cite{amendola_cosmology_2016, laureijs_euclid_2011}, complemented with the expected CMB information from Simons Observatory \cite{the_simons_observatory_collaboration_simons_2019}, but discuss the changes in the results for other surveys in Section \ref{subsec_alternative_galaxy_surveys}. In Section \ref{sec_discussion} we provide some discussion and caveats, along with comparisons to previous work in the literature. We conclude in Section \ref{sec_conclusions}. 

\section{Methodology}\label{sec_methodology}

\subsection{Implementation of the NLO Power Spectra}

One of the goals of this work is to compare the results obtained when performing forecasts using the linear and next-to-leading-order (NLO) or 1-loop galaxy power spectrum. The Kaiser formula describes the linear redshift-space galaxy power spectrum at redshift $z$:

\begin{equation}\label{eq_lin_pgg}
P^{lin.}_{gg,s}(k,z) = [b_1(z)+f_{bc}(k,z)\mu^2]^2P_{bc}(k,z)+ P_\epsilon^{\{0\}}(z).
\end{equation}
Here, $P_\epsilon^{\{0\}}$ is the stochastic (shot noise) contribution,
$P_{bc}(k,z)$ is the linear baryon and cold dark matter power spectrum (obtained using the Boltzmann code \texttt{CLASS} \cite{blas_cosmic_2011}), while $f_{bc}(k,z) \equiv d\ln D(k,z)/d\ln a$ is the scale-dependent linear growth rate. It has been verified by simulations \cite[e.g.][]{castorina_demnuni_2015} that because neutrinos do not cluster on the scales on which galaxies form, the baryon and cold dark matter only power spectrum $P_{bc}(k,z)$ should be used in the calculation of the galaxy power spectrum, along with its corresponding growth rate, $f_{bc}(k,z)$. The presence of massive neutrinos induces a scale-dependence in $f_{bc}(k,z)$ and this contributes to a small degree to constraints on $M_\nu$ \cite{hernandez_neutrino_2017, boyle_deconstructing_2018} and is therefore included in Equation \ref{eq_lin_pgg}. In the following, we will drop the redshift argument for clarity, noting that all bias parameters, power spectra, as well as the growth rate $f_{bc}$ are to be evaluated at the redshift of observation.

The NLO redshift-space power spectrum (which includes the linear component above) provided by \cite{desjacques_galaxy_2019} is calculated using a total of 28 independent loop integrals, and requires specification of 5 additional free bias parameters, 1 velocity bias parameter and 3 stochastic parameters if selection effects are neglected, as they are throughout most of this work. We note that the effects of massive neutrinos were not explicitly considered in \cite{desjacques_galaxy_2019}, and our results are therefore based on the assumption that the important effects of massive neutrinos are captured by their contribution to the linear power spectrum and growth factor (specifically of the CDM and baryon components only, as described above). This approximation was first introduced in \cite{saito_impact_2008, saito_nonlinear_2009}. In Section \ref{subsec_neutrino_bias}, we discuss the choice to neglect the specific effect of massive neutrinos on the bias parameters. The NLO redshift-space galaxy power spectrum can be summarised in the following form at a specific redshift:

\begin{equation}\label{eq_pk_nl}
P_{gg,s}^{\textrm{NLO}}(k,\mu)=P_{gg,s}^{\textrm{lb+hd}}+P_{gg,s}^{2-2}(k,\mu)+2P_{gg,s}^{1-3}(k,\mu)
\end{equation}

\begin{equation}\label{eq_nl_pl_hd}
\begin{split}
P_{gg,s}^{\textrm{l+hd}}(k,\mu) =~&[b_1+f_{bc}(k)\mu^2]^2 P_{bc}(k)+P_\epsilon^{\{0\}} \\& -2\{b_1\blapl+\mu^2f_{bc}(k)[\blapl+b_1\beta_{\nabla^2v}]\\& \qquad +\mu^4f_{bc}^2(k)\beta_{\nabla^2v}\}k^2P_{bc}(k)\\&+k^2P_\epsilon^{\{2\}}-\mu^2k^2P_{\epsilon\varepsilon_\eta}^{\{2\}}
\end{split}
\end{equation}

%\begin{equation}\label{eq_nl_pl_hd}
%\begin{split}
%P_{gg,s}^{l+hd}(k,\mu) =~&[b_1-b_\eta f\mu^2]^2 P_L(k)+P_\epsilon^{\{0\}} \\& -2\{b_1b_{\nabla^2\delta}-\mu^2fb_\eta[b_{\nabla^2\delta}+b_1\beta_{\nabla^2v}+b_1\beta_{\delta_\parallel^2 v}\mu^2]\\& +\mu^4f^2b_\eta^2[\beta_{\nabla^2v}+\beta_{\delta_\parallel^2 v}\mu^2]\}k^2P_L(k)\\&+k^2P_\epsilon^{\{2\}}-\mu^2k^2P_{\epsilon\varepsilon_\eta}^{\{2\}},
%\end{split}
%\end{equation}

\begin{equation}
P_{gg,s}^{2-2}(k,\mu) = \sum_{n=0}^4\sum_{(m,p)}A_{n(m,p)}(f_{bc}(k),b_{\mathcal{O}})\mathcal{I}_{mp}(k)\mu^{2n}
\end{equation}

\begin{equation}
P_{gg,s}^{1-3}(k,\mu) = \sum_{l=0,2,4,6}\sum_{n=1}^5 C_n^{1-3,l}(f_{bc}(k),b_\mathcal{O})\mathcal{I}_n(k)P_{bc}(k)\mathcal{L}_{l}(\mu)
\end{equation}

In Eq. \ref{eq_nl_pl_hd} we have defined the parameter combination $\blapl = C_s^2 + b_{\nabla^2\delta}$ of the higher-derivative bias proper and the effective sound speed $C_s^2$. The distinction between $\blapl$ and $b_{\nabla^2\delta}$ will become relevant when also including the galaxy-matter cross-power spectrum as probed via the cross-correlation with CMB lensing.

The form of the 23 $\mathcal{I}_{mp}$ and 5 $\mathcal{I}_{n}$ loop integrals are provided in \cite{desjacques_galaxy_2019}, and the coefficients $A_{n(m,p)}$ and $C_n^{1-3,l}$ and their input bias parameters $b_\mathcal{O}$ can be obtained from a \texttt{MATHEMATICA} notebook provided as supplementary material\footnote{\label{footnote:github}\url{https://github.com/djeong98/pkgs_supplement}}. These coefficients significantly improve the calculation time by combining contributions to give the minimal number of integrals. We highlight again that we include the scale-dependence in $f_{bc}(k)$ when performing our calculations. 

%\chg{\sout{Table \ref{table_fiducial_bias} shows the 9 free nuisance parameters that remain if selection effects are neglected and how their fiducial values were calculated. $R_L$ represents the Lagrangian radii of haloes, calculated using approximate typical halo masses at each redshift derived from the halo mass function of \cite{tinker_toward_2008}. $n_{g}$ is the galaxy number density. The choice for $b_1$ was made for consistency with \cite{boyle_deconstructing_2018, boyle_understanding_2019}. The fitting formula for $b_2$ was taken from N-body simulations and provided by \cite{lazeyras_precision_2016}. The values of $b_{K^2}$ and $b_{td}$ were both derived from the Lagrangian local (in the matter density) assumption \cite{desjacques_galaxy_2019}.}} \chg{\textbf{[Paragraph moved later.]}}

We also analyse the constraining power of CMB lensing. The CMB lensing power spectrum is calculated using the Limber approximation as

\begin{equation}\label{eq_cl_kk}
C_l^{\kappa\kappa} = \left(\frac{4\pi G\rho_{m,0}}{c^2}\right)^2 \int_0^{z_{\star}}dz (1+z)^2 \left(\frac{d_{A}(z,z_{\star})}{d_{A}(z_{\star})}\right)^2 \frac{P_{mm}^{NLO}\left[k=\frac{l+1/2}{d_{A}(z)},z\right]}{H(z)}.
\end{equation}

$\rho_{m,0}$ is the comoving matter density, $z_{\star}$ is the redshift of last scattering, and $d_{A}(z,z_{\star})$ represents the comoving angular diameter distance between the two redshifts. $P_{mm}^{NLO}$ is the NLO matter power spectrum (here including the mass contributed by neutrinos)\. It is calculated as $P_{mm}^{NLO} = P_L(k) + P_{22}(k) + 2 P_{13}(k) - 2 C_s^2 k^2 P_L(k)$, with $P_{22}(k)$ and $P_{13}(k)$ being calculated using a publicly available code\footnote{https://wwwmpa.mpa-garching.mpg.de/\~{}komatsu/crl/list-of-routines.html} \cite{jeong_perturbation_2006}. Once again, we make the assumption that the effect of massive neutrinos is sufficiently well accounted for by including only the contribution of neutrino perturbations to the linear power spectrum. Calculating the matter power spectrum at the large number of redshifts required by Equation \ref{eq_cl_kk} would be very time-consuming, but it is also improper to scale between different redshifts using the growth factor $D(z)$ because of the redshift-dependent shape of the power spectrum when massive neutrinos are included. To account for this, we calculate the NLO matter power spectra independently for $0<z<3$, but scale backwards using the growth factor beyond $z=3$, where the suppression due to free-streaming is less substantial. 

Finally, we can also consider the cross-correlation between galaxy clustering and CMB lensing as a probe. The galaxy-CMB lensing power spectrum is calculated (also using the Limber approximation) as

\begin{equation}\label{eq_cl_gk}
C_l^{g\kappa} = \left(\frac{4\pi G\rho_{m,0}}{c^2}\right)(1+z_g)\frac{d_{A}(z_g,z_{\star})}{d_{A}(z_g)d_{A}(z_{\star})}P_{gm}\left[k=\frac{l+1/2}{d_{A}(z_g)},z_g\right],
\end{equation}

The NLO galaxy-matter cross-power spectrum required in Equation \ref{eq_cl_gk} can be derived straightforwardly from the NLO galaxy power spectrum presented in \cite{desjacques_galaxy_2019}. As we use the Limber approximation, we perform the calculation for $\mu=0$ only. The components can be calculated as follows:

\begin{equation}\label{eq_nl_pl_hd_mu0}
\begin{split}
P_{gm,s}^{\textrm{l+hd}}(k) = b_1P_L(k)+ ( b_1 C_s^2 - \blapl) k^2P_L(k) + k^2P_{\epsilon,gm}^{\{2\}}
\end{split}
\end{equation}

\begin{equation}
P_{gm,s}^{2-2}(k) = \sum_{(m,p)}A_{cross(m,p)}(f_{bc}(k),b_{\mathcal{O}})\mathcal{I}_{mp}(k)
\end{equation}

\begin{equation}
P_{gm,s}^{1-3}(k) = \sum_{l=0,2,4,6}\sum_{n=1}^5 C_{cross,n}^{1-3,l}(f_{bc}(k),b_\mathcal{O})\mathcal{I}_n(k)P_{L}(k)\mathcal{L}_{l}(0)
\end{equation}

$A_{cross(m,p)}$ has 19 non-zero terms and $C_{cross,n}^{1-3,l}$ has 12 without selection effects. These coefficients are available in a Mathematica notebook alongside the supplementary material\textsuperscript{\ref{footnote:github}} provided in \cite{desjacques_galaxy_2019}. For the effective matter sound speed $C_s^2$, we take the value determined from N-body simulations in \cite{Lazeyras_2019} at redshift zero of 1.31 $h^{-2}~{\rm Mpc}^2$, and rescale with redshift as $C_s^2(z)=[D(z)/D(0)]^4C_s^2(0)$, which was found to describe the redshift dependence of the effective sound speed well. For the fiducial value of the cross-stochasticity between matter and galaxies, $P_{\epsilon,gm}^{\{2\}}$, for which no reliable simulation measurements exist to our knowledge,  we take $C_s^2P_{\epsilon}^{\{0\}}$. We stress however, that both $C_s^2$ and $P_{\epsilon,hm}^{\{2\}}$  have an entirely negligible effect on our results.

We perform all of our calculations using the Fisher matrix formalism (see \cite{boyle_deconstructing_2018} for details).
Our code has been thoroughly validated through comparisons of results with those of \cite{agarwal_information_2020}. 
The covariance for the three-dimensional galaxy power spectrum is calculated as: 

\begin{equation}\label{eq_covariance_3D}
\langle \Delta P_{gg,s}(k,\mu)^{2}\rangle = \frac{2\pi^2}{Vk^2\Delta k\Delta\mu}2\left[P_{gg,s}(k,\mu)\right]^2,
\end{equation}
where $V$ is the volume of the redshift bin being observed, and $\Delta k$ and $\Delta \mu$ are the bin sizes for the wavenumber and angle with respect to the line of sight, respectively. The covariances for the angular power spectra are calculated as:

\begin{equation}\label{eq_covariance_2D}
\langle \Delta C_{l}^{xy} \Delta C_{l}^{mn} \rangle = \frac{1}{(2l+1)f_{\textrm{sky}}\Delta l}(C_{l}^{xm}C_{l}^{yn} + C_{l}^{xn}C_{l}^{ym}).
\end{equation}

Here, $f_{\textrm{sky}}$ is the fraction of the sky observed. The $C_{l}$ values on the right-hand side of Equation \ref{eq_covariance_2D} must include appropriate noise terms for auto-correlation power spectra.

\subsection{Fiducial Cosmology, Surveys and Priors}\label{subsec_fiducials}

\begin{table}
\centering
\begin{tabular}{l | l}
  Bias parameter & Fiducial value\\
  \hline
$b_1$ & $b_1(z)D(z)=0.76$\\
$\blapl$ & $R_{L}^2$\\
$\beta_{\nabla^2v}$ & $R_{L}^2$\\
$b_2$ & $0.412-2.143b_1+0.929(b_1)^2+0.008(b_1)^3$\\
$b_{K^2}$ & $-\frac{2}{7}(b_1-1)$\\
$b_\textrm{td}$ & $\frac{23}{42}(b_1-1)$\\
$P_\epsilon^{\{0\}}$ & $n_{g}^{-1}$\\
$P_\epsilon^{\{2\}}$ & $0$\\
$P_{\epsilon\varepsilon_{\eta}}^{\{0\}}$ & $0$\\
\end{tabular}
\caption{Fiducial values assumed for the 6 bias parameters and 3 stochastic parameters required for calculation of the NLO galaxy power spectrum. $R_L$ denotes the Lagrangian radius of halos relevant at each redshift (see Appendix \ref{app_surveys}). Note that $\blapl$ also includes the effective sound speed for matter, but this is a subdominant contribution.
}\label{table_fiducial_bias}
\end{table}

In \cite{boyle_deconstructing_2018} we used Euclid \cite{amendola_cosmology_2016, laureijs_euclid_2011} as a sample survey for which to forecast constraints. In \cite{boyle_understanding_2019}, we extended our work to include future CMB surveys and focused on Simons Observatory \cite{the_simons_observatory_collaboration_simons_2019} as our example. Here we continue to focus on these two as representative surveys. The survey parameters for Euclid are given in Table \ref{table_euclid} in Appendix \ref{app_surveys}, and those for Simons Observatory are taken from Table 1 of \cite{the_simons_observatory_collaboration_simons_2019}. The Euclid survey area is 15000 deg$^2$ and the total number of galaxies included in the spectroscopic survey is approximately 50 million. The focus of our work is not to provide quantitative information for a particular survey, but to demonstrate certain principles, and these should be generally extendable to all similar datasets. Results for alternative galaxy surveys are discussed in Section \ref{subsec_alternative_galaxy_surveys}. We assume a maximum usable wavenumber of $k_{\textrm{max}} = 0.2~h~\textrm{Mpc}^{-1}$ for all galaxy survey calculations for ease of comparison with results in our previous papers, and the effect of this choice is examined in Section \ref{subsec_fisher_bias}. We calculate a consistent $l_{\rm max}$ value in each redshift bin for the galaxy-CMB cross-correlation calculations (Eq. \ref{eq_cl_gk}) as $l_{\rm max}(z)=k_{\rm max}\times d_A(z)-1/2$. For the CMB lensing auto-correlation (Eq. \ref{eq_cl_kk}), no additional $l_{\rm max}$ is implemented because of the natural limit imposed by the lensing noise. We have checked that repeating our CMB lensing calculations with full Halofit power spectra (as implemented in \texttt{CLASS}) has negligible impact on the results.

Our fiducial cosmology is identical to that in \cite{boyle_deconstructing_2018, boyle_understanding_2019}. We have a total of 11 free cosmological parameters. We marginalise our constraints on $M_\nu$ over the six standard Planck convention $\Lambda$CDM parameters ($\theta_s$, $\omega_b$, $\omega_{cdm}$, $\tau$, $A_s$ and $n_s$) plus $N_{\textrm{\scriptsize{eff}}}$. In all cases, we examine the effects of freeing $\Omega_{k}$, $w_0$ and $w_a$ on the constraints.

Table \ref{table_fiducial_bias} shows the 9 free nuisance parameters that remain if selection effects are neglected and how their fiducial values were calculated. $R_L$ represents the Lagrangian radii of haloes, calculated using approximate typical halo masses at each redshift derived from the halo mass function of \cite{tinker_toward_2008} based on an abundance-matching procedure. Specifically, we use the median mass of halos above a threshold adjusted to yield the expected number density of galaxies $n_g$ in each redshift bin as representative halo mass. The fiducial values of $C_{\nabla^2\delta}$ and $\beta_{\nabla^2\delta}$ are set to $R_L^2$ based on dimensional reasoning. We do not expect the precise values of the fiducial higher-derivative bias parameters to have any impact on our results. The choice for $b_1$ was made for consistency with \cite{boyle_deconstructing_2018, boyle_understanding_2019}. The fitting formula for $b_2$ was taken from N-body simulations and provided by \cite{lazeyras_precision_2016}. The values of $b_{K^2}$ and $b_{td}$ were both derived from the Lagrangian local (in the matter density) assumption \cite{desjacques_galaxy_2019}.

Our basic CMB prior is also the same as that used in \cite{boyle_understanding_2019}. We combine the temperature power spectrum from Planck \cite{planck_collaboration_planck_2019}, a forecasted E-mode polarisation power spectrum from Simons Observatory and the cross-correlation between them. These are the unlensed spectra, as we choose to analyse the impact of including CMB lensing separately from that of the primary anisotropies. Note that when this is not of interest, one should perform these calculations using the lensed temperature and polarisation power spectra, as this is what is measured in real experiments. Delensing of measured spectra is a complicated process, and it has been demonstrated previously that the cross-correlation between lensed temperature/polarisation power spectra and the lensing power spectrum is very small \cite{peloton_full_2017, rocher_probing_2007}, so the results will not be misleading. We also include a prior on $\sigma(\tau)=0.008$ from Planck (as given for TT,TE,EE+lowE in \cite{planck_collaboration_planck_2019}). 

\subsection{Constraining Methods}\label{subsec_constraining_methods}

We perform forecasts in this article using three different approaches to the galaxy power spectrum. The most optimistic cosmological parameter constraint forecasts for a particular galaxy survey can be extracted by using the full galaxy power spectrum given in Equation \ref{eq_pk_nl} as the observable with which derivatives for the Fisher matrix are calculated. To include full geometric information, the power spectrum must be adjusted to account for the uncertainty in $H(z)$ and $D_{A}(z)$ when converting observed scales to physical comoving scales, as follows:

\begin{equation}
P(k_{\parallel}^{\textrm{obs}}, k_{\perp}^{\textrm{obs}}) =\frac{H(z)}{H_{fid}(z)}\left(\frac{D_{A, fid}(z)}{D_{A}(z)}\right)^2 P(k_{\parallel}^{\textrm{com}}, k_{\perp}^{\textrm{com}}),
\end{equation}
where $k_{\parallel}^{\textrm{obs}} = k_{\parallel}^{\textrm{com}}(H_{fid}(z)/H(z))$ and $k_{\perp}^{\textrm{obs}} = k_{\perp}^{\textrm{com}}(D_A(z)/D_{A,fid}(z))$. These \lq combined\rq~constraints inherently combine information from many different sources within the power spectrum - constraints on distances from characteristic scales (e.g. BAO) and from the Alcock-Paczynski test, constraints on the structure growth rate as a function of redshift from redshift space distortions, and constraints from the overall shape and amplitude of the power spectrum itself. As shown in \cite{boyle_deconstructing_2018}, the constraints extracted in this way can depend heavily on the underlying cosmology, degrading significantly when extensions like non-zero curvature or a varying dark energy equation of state are added. 

However, our work in \cite{boyle_understanding_2019, boyle_deconstructing_2018} demonstrated that robust, cosmology-independent constraints on the neutrino mass can be extracted from the galaxy power spectrum by isolating a particular probe. Massive neutrinos suppress the growth of structure on small scales relative to that on large scales to an extent that is primarily dependent on the total mass of the neutrino species by free-streaming out of potential wells below a characteristic scale. This process imprints a characteristic scale-dependent signature in the shape of the matter power spectrum and in the structure growth rate. Because changes in other cosmological parameters cannot replicate this effect, the constraints extracted from this scale-dependent suppression in the power spectrum and structure growth rate (we call these \lq free-streaming\rq~constraints) are independent of the assumed cosmology, and therefore give much more reliable constraints than those extracted from the full combined power spectrum. 

The details of how the free-streaming constraints are achieved in the linear case are given in \cite{boyle_deconstructing_2018}. In the NLO case, we normalise the input $P_{bc}(k)$ and $f_{bc}(k)$ at a fixed scale before calculating the components of Equation \ref{eq_pk_nl} and then calculating numerical derivatives. Therefore only the scale-dependence in the input $P_{bc}(k)$ and $f_{bc}(k)$ contributes. Smoothing is also applied to remove any wiggles in the derivatives due to changes in the BAO signal. We marginalise the final Fisher matrix over the uncertainty in the amplitudes of $P_{bc}(k)$ and $f_{bc}(k)$. The free-streaming information can be combined with corresponding signatures in the CMB lensing and galaxy-CMB lensing power spectra by marginalising over the large-scale amplitudes of both. 

Finally, we present constraints using distance information only extracted from the isolated BAO signal (fitted from the galaxy power spectrum). This proceeds in the same manner as outlined in \cite{boyle_deconstructing_2018}. A two-step Fisher matrix process is used to first constrain $H(z)$ and $D_A(z)$, then the cosmological parameters. For these constraints (but not for the combined constraints) we also implement an IR resummation procedure in our model to account for BAO damping. In Section \ref{subsec_bao} we show that the impact of this on the constraints on $M_\nu$ is small enough to be excluded in the case of the combined constraints. 

\section{Results}\label{sec_results}

\subsection{Combined Constraints}

\begin{figure}
\includegraphics[width=\textwidth]{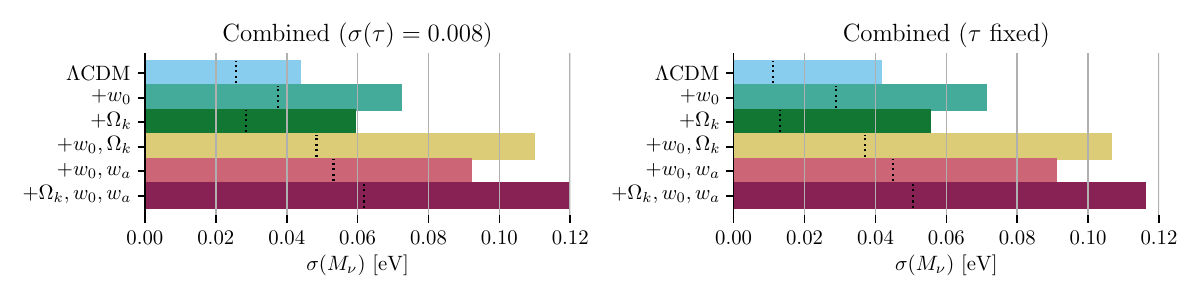}
\caption{Forecasted 1$\sigma$ constraints on $M_\nu$ for various cosmological models, using the full NLO galaxy power spectrum for Euclid. This represents the most optimistic forecast and inherently includes BAO, RSD and shape information. The CMB prior combines the temperature power spectrum from Planck with a forecasted polarisation power spectrum from Simons Observatory (see Section \ref{subsec_fiducials}). The left panel shows the constraints with a $1\sigma$ prior on $\tau$ of 0.008 (from Planck 2018 \cite{planck_collaboration_planck_2019}), and the right panel shows how the constraints would improve if $\tau$ were perfectly constrained through some external experiment. In both cases, the dotted tick marks show the constraints if linear theory was asssumed for the galaxy power spectrum (with the same $\kmax$ of 0.2 $h~\textrm{Mpc}^{-1}$). Note that CMB lensing is neglected here but included in Figure \ref{fig_combined_cmb_lensing}.}\label{fig_combined}
\end{figure}

\begin{figure}
\centering
\includegraphics[width=0.75\textwidth]{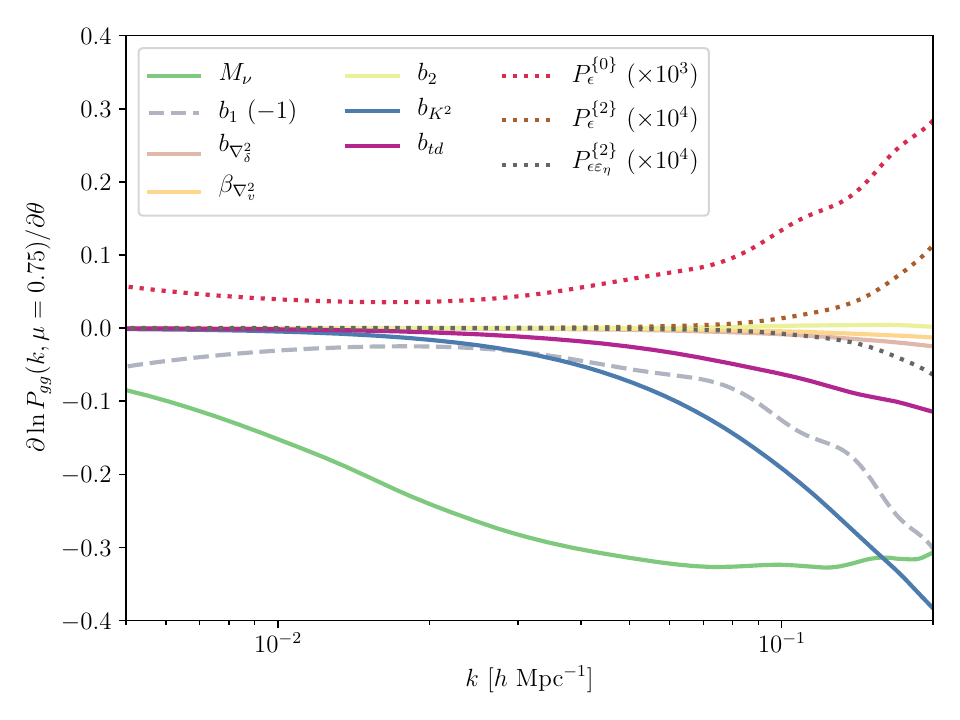}
\caption{The log derivatives of the NLO galaxy power spectrum (for $\mu=0.75$) with respect to $M_\nu$ and the nine nuisance parameters at $z=1.35$ (the central redshift of Euclid). A contant factor of 1.0 has been subtracted in the $b_1$ case and the stochastic parameter derivatives have been rescaled by a multiplicative factor (see the legend) for the readability of the plot.}\label{fig_derivatives}
\end{figure}

\begin{figure}
\makebox[\textwidth][c]{\includegraphics[width=1.2\textwidth]{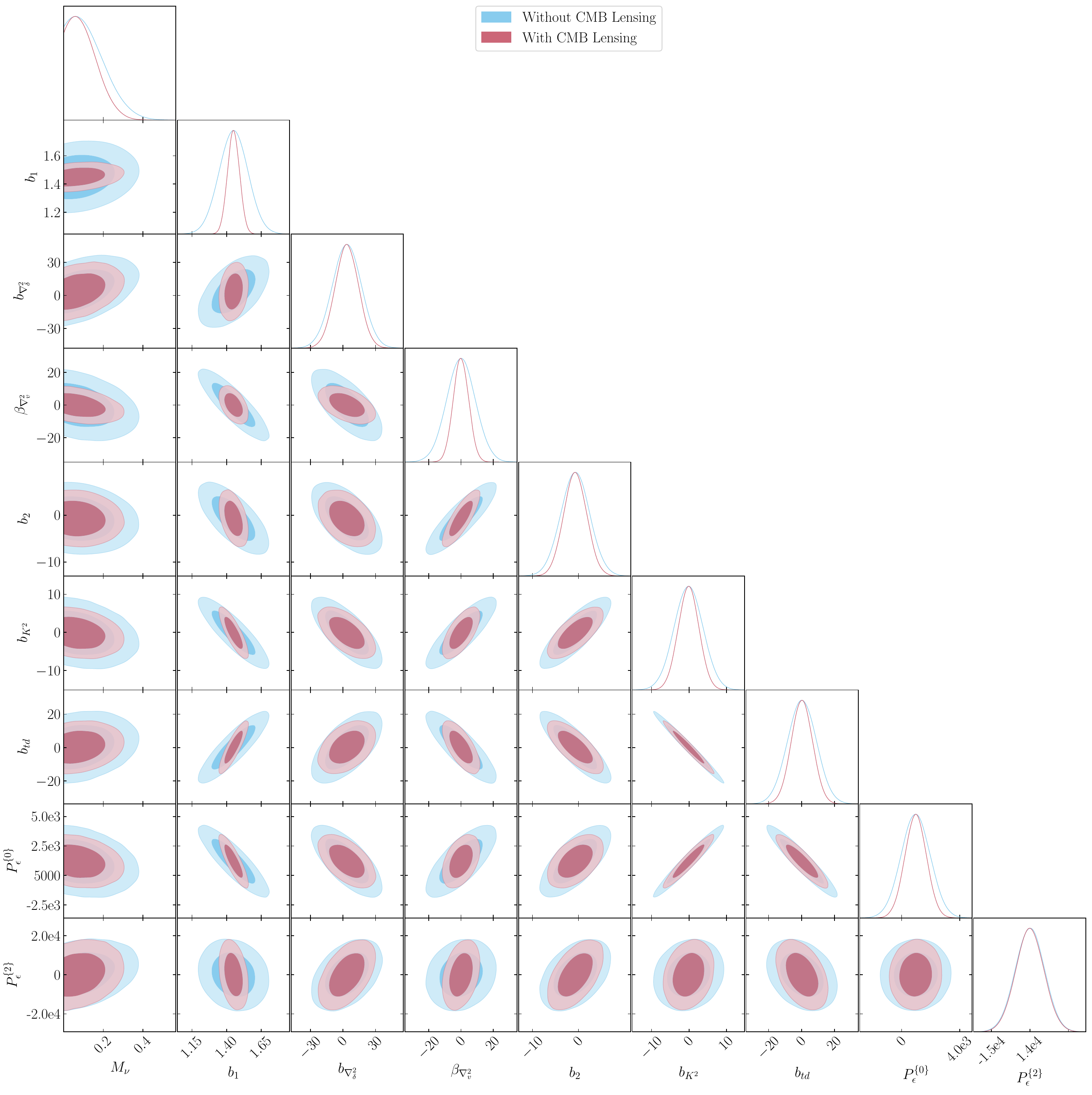}}
\caption{Contours showing the relationship between the different nuisance parameters and $M_\nu$ for the central redshift bin of our Euclid model plus a CMB prior. The blue contours represent the \lq combined\rq~galaxy clustering case without CMB lensing and the pink contours represent the case when CMB lensing is added, reducing the dependence of the constraints on the nuisance parameters. The parameter $P_{\epsilon\varepsilon_{\eta}}^{\{0\}}$ is excluded because of its negligible impact.
}\label{fig_contour}
\end{figure}

The left panel of Figure \ref{fig_combined} shows the forecasted constraints on $M_\nu$ for Euclid using the full galaxy power spectrum as a constraining tool, along with the priors outlined in Section \ref{subsec_fiducials}. The vertical dotted tick marks show the constraints predicted if the tree-level galaxy power spectrum is assumed instead. We see that the constraints are significantly weakened on updating our results with the NLO power spectrum. As found for the linear galaxy power spectrum in our previous work \cite{boyle_deconstructing_2018, boyle_understanding_2019}, these combined constraints are strongly cosmology-dependent, and constraints assuming flatness or a simple cosmological constant should therefore be presented with appropriate caveats both in forecasts and when derived from real data. 

The right panel shows the same constraints as the left panel but with $\tau$ held fixed. It is interesting to note that the $M_\nu$-$\tau$ degeneracy, a significant focus of much recent forecast work on neutrino mass constraints from large-scale structure \cite[e.g.][]{allison_towards_2015, boyle_deconstructing_2018, boyle_understanding_2019, mishra-sharma_neutrino_2018, yu_towards_2018, brinckmann_promising_2019, archidiacono_physical_2017} becomes much less significant when considering the NLO galaxy power spectrum (note the change in the linear galaxy power spectrum constraints between the two panels). The $M_\nu$-$\tau$ degeneracy arises as a result of the combination of primary CMB and large-scale structure data. Measurements of the CMB temperature power spectrum constrain the parameter combination $A_s\exp(-2\tau)$. Massive neutrinos affect the growth of structure and therefore $M_\nu$ becomes correlated with $A_s$ in large-scale structure measurements. When CMB and large-scale structure data are combined, the result is a strong degeneracy between $M_\nu$ and $\tau$. In the right panel of Figure \ref{fig_combined}, however, we see that this plays a much lesser role in the NLO case, as the neutrino mass constraint is supported primarily by other degeneracies. 

Figure \ref{fig_derivatives} shows the log derivatives of $P_{gg,s}$ with respect to $M_\nu$ and the 9 nuisance parameters. We see that several of these parameters, in particular nonlinear bias parameters, change the power on small scales, and therefore lessen the information on $M_\nu$ available from its characteristic suppression of the power spectrum on these scales. Using the NLO power spectrum but fixing the additional free parameters recovers almost the same constraints as in the linear power spectrum case (0.026 eV) for $\Lambda$CDM, while the constraint weakens to 0.044 eV when the new parameters are left free to vary. All of the nuisance parameters are somewhat correlated with $M_\nu$ (except $P_{\epsilon\varepsilon_{\eta}}^{\{0\}}$, which has little effect), and of particular importance is that the nuisance parameters are quite strongly correlated with each other (see Figure \ref{fig_contour}). 

In the linear case, imposing a prior on $b_1$ has the potential to significantly improve constraints on $M_\nu$: if $b_1$ is fixed in each redshift bin, the $M_\nu$ constraint falls from 0.026 eV to 0.021 eV. In the NLO case, fixing $b_1$ also improves the constraint, from 0.044 eV to 0.034 eV. Without any prior on $b_1$, the constraint in a typical redshift bin is about a factor of 7 weaker in the NLO case compared to in the linear case, because of the degeneracies with other nuisance parameters. 

If $b_1$ is well constrained, adding priors on any of the other bias parameters provides further improvements. Fixing the two bias parameters associated with operators derived from the tidal field, $b_{K^2}$ and $b_{td}$, provides a further improvement in the constraints on $M_\nu$ to 0.031 eV. $M_\nu$ shows some correlation with both parameters. However, these two parameters are very strongly anti-correlated with each other, as is clear from Figure \ref{fig_contour}. We discuss the reason for this strong degeneracy in Section \ref{subsec_bias_degeneracy}. $M_\nu$ shows some correlation with both parameters. Because of the strong degeneracy between the parameters, fixing one without the other has very little impact on the neutrino mass constraint, as the other can compensate for its effect. In this way, these two bias parameters behave like a single effective parameter in this case. $P_\epsilon^{\{0\}}$ is also strongly correlated with $b_{K^2}$. 

Fixing $b_1$ and the two tidal bias parameters improves the constraint to within 20\% of that in the linear case. Fixing, for example, $b_1$, $\blapl$ and $\beta_{\nabla^2v}$ achieves a similar constraint. Because the bias parameters are so correlated, adding priors on a few of them has a significant effect on the constraints on the others and therefore on the constraint on $M_\nu$. Robust theoretical priors on the NLO bias parameters could potentially have a significant impact in closing the gap between these results and those derived for the linear power spectrum. Alternatively, measurements of the bispectrum could help reduce the uncertainty on $b_1$.

\begin{figure}
\centering
\includegraphics[width=\textwidth]{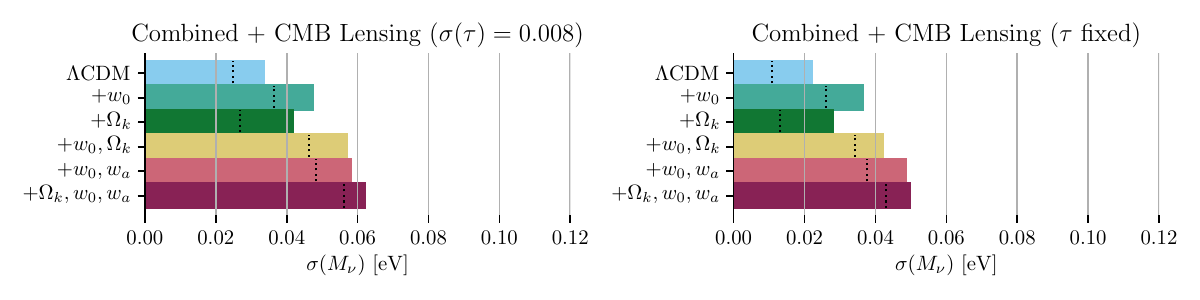}
\caption{The NLO combined constraints with and without CMB lensing. The cross correlation between galaxies and CMB lensing is also included but has very little effect. The same results for the tree-level galaxy power spectrum and linear CMB lensing power spectrum are shown as the vertical tick marks. Comparing to Figure \ref{fig_combined}, one sees that CMB lensing plays a much more substantial role when combined with the 1-loop power spectrum, by helping to break degeneracies between the cosmological parameters and the large number of nuisance parameters. Once these degeneracies are under control, we see in the right panel that the optical depth to reionisation $\tau$ becomes a significant driver of the neutrino mass constraint once again.}\label{fig_combined_cmb_lensing}
\end{figure}

In \cite{boyle_understanding_2019}, we showed that future CMB lensing data will add little to the combined constraints on the neutrino mass from large-scale structure measurements. However, as the NLO constraints are weakened, it is likely that the relative contribution from CMB lensing will be more substantial. The CMB lensing power spectrum is a function of the total matter power spectrum (including neutrino perturbations) and therefore measurements are unencumbered by the new galaxy power spectrum parameters introduced. 

Figure \ref{fig_combined_cmb_lensing} shows how the addition of forecast CMB lensing data from Simons Observatory could improve the NLO constraints. CMB lensing contributes primarily by breaking degeneracies between the nuisance parameters and the cosmological parameters (see also Figure \ref{fig_contour}). It also reduces the cosmology dependence of the constraints. The galaxy-CMB lensing cross power spectrum is also included here, but makes very little contribution. Interestingly, once information from CMB lensing is added and the degeneracies between $M_\nu$ and the bias parameters are under control, the degeneracy with $\tau$ once again becomes a significant factor in the overall constraints. CMB lensing contributes in particular by improving constraints on $A_s$, which in turn improves constraints on $b_1$, which also impacts the large-scale amplitude of the power spectrum. 

\subsection{Free-Streaming Constraints}\label{subsec_free_streaming}

\begin{figure}
\centering
\includegraphics[width=0.525\textwidth]{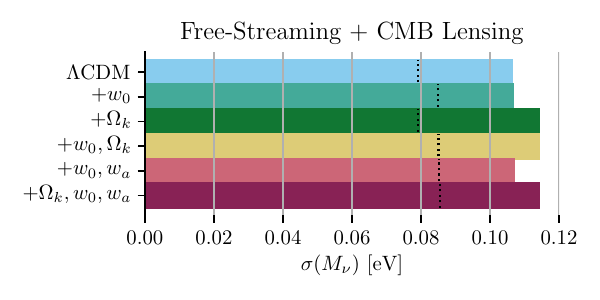}
\caption{As Figure \ref{fig_combined}, but with the constraints derived using only the scale-dependent free-streaming signals in the galaxy, CMB lensing and galaxy-CMB lensing power spectra as a probe (see Section \ref{subsec_constraining_methods}). One can see that the constraints are quite independent of all the cosmological extensions considered in both the linear and NLO cases. These constraints are independent of $\tau$.}\label{fig_fs}
\end{figure}

\begin{figure}
\centering
\includegraphics[width=0.65\textwidth]{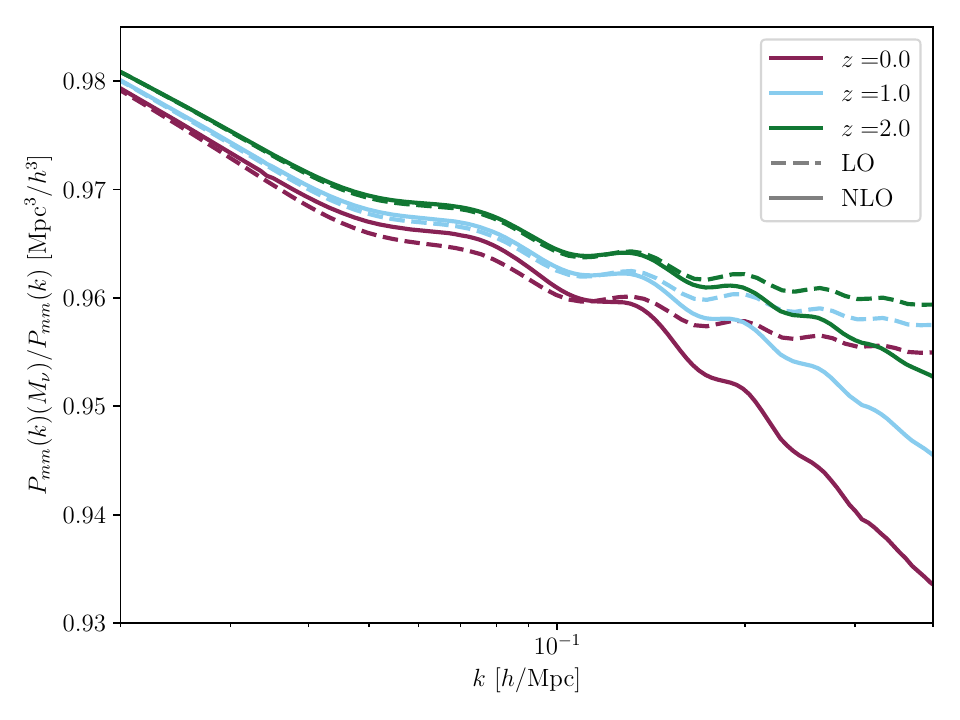}
\caption{The suppression in the matter power spectrum caused by the addition of a single massive neutrino of mass 0.06 eV relative to the case with no massive neutrinos, for both the linear and 1-loop/NLO cases. The relative suppression is enhanced in the NLO case. This enhanced effect propagates into both the galaxy and CMB lensing power spectra.}\label{fig_suppression}
\end{figure}

Figure \ref{fig_fs} shows the forecasted constraints on the total neutrino mass using only the scale-dependence of the power spectrum as a constraining tool (see Section \ref{subsec_constraining_methods}). It can be seen that although the constraints are weakened somewhat by updating to the NLO power spectrum, they remain independent of changes in curvature and the dark energy equation of state. However, they degrade from around 0.08 eV to approximately 0.11 eV. This is a smaller relative change than in the combined case.

Interestingly, fixing all of the new nuisance parameters in the NLO case improves the constraint only to around 0.1 eV. From Figure \ref{fig_suppression}, we can see that the free-streaming signal is actually enhanced in the NLO case (Figure \ref{fig_suppression} shows this for the matter power spectrum), and on investigation of the Fisher matrix, the information content on $M_\nu$ is actually increased. So the cause of this difference is not a change in the strength of signal in the power spectrum itself. The issue lies in the new degeneracy with $b_1$. Note that in the case of the linear power spectrum, extracting only the scale-dependence means treating the amplitude of the power spectrum as a free parameter, and $b_1$ is a simple multiplicative factor, so has no effect on the final constraint on $M_\nu$. However, in the NLO case $b_1$ now has a scale-dependent component and therefore some degeneracy with $M_\nu$. Fixing $b_1$ plus all the new nuisance parameters gives a slightly better constraint in the NLO case than in the linear case, as expected. 

One can now ask which of the new parameters contribute in particular to the weakening of the constraints on $M_\nu$ seen in Figure \ref{fig_fs}. The story is generally very similar to that in the combined case, but with significant changes for $b_1$ and $P_\epsilon^{\{0\}}$. This makes sense, as the other nuisance parameters do not change the large-scale power spectrum, so their effects when only the scale-dependence is considered are identical to those in the combined case.

Ultimately, considering that seven new nuisance parameters have been added to the calculation, all with scale-dependent effects, it could be considered surprising that a degradation in the overall constraints of only about 40\% is observed. The cause of this is that the scale-dependence caused by neutrino free-streaming begins to take effect on much larger scales than those on which the scale-dependence of the bias parameters starts to be significant. Increasing the total neutrino mass shifts the free-streaming scale to higher $k$, and could therefore lead to greater degeneracy with the new parameters. This possibility is explored in Section \ref{subsec_higher_mnu}. 

\subsection{BAO Constraints}\label{subsec_bao}

\begin{figure}
\centering
\includegraphics[width=0.525\textwidth]{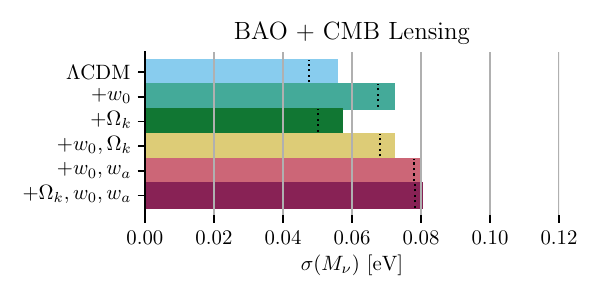}
\caption{As Figure \ref{fig_fs}, but with the constraints derived from distance information from the BAO signal only.}\label{fig_bao}
\end{figure}

Figure \ref{fig_bao} shows the forecasted constraints on $M_\nu$ when only constraints on the expansion rate $H(z)$ and angular diameter distance $D_{A}(z)$ provided by the baryon acoustic oscillations (BAO) in the power spectrum are used as a constraining tool. Importantly, CMB lensing is included, which has a massive impact on breaking cosmological degeneracies in this case. As is well-known, BAO are favoured for their robustness against non-linear effects, and that is reflected here. We see that the constraints degrade little when the 1-loop power spectrum is used instead of the linear power spectrum compared to the case of the full galaxy power spectrum. The constraint on $M_\nu$ is degraded by approximately 20\% in the $\Lambda$CDM case, and much less so for the more complex cosmologies, where the relative contribution of BAO measurements to the overall constraint is diminished. As in the linear case, the constraints remain cosmology-dependent, with a particular sensitivity to the dark energy equation of state (the curvature degeneracy is alleviated by the inclusion of CMB lensing).

We emphasise that in the calculation for Figure \ref{fig_bao}, we include BAO damping using an IR-resummation procedure \cite{senatore/zaldarriaga:2015,baldauf/etal:2015BAO,blas/etal:2016,Perko:2016puo,senatore/trevisan:2018}, as described below, although BAO damping is not included in our main model of the NLO galaxy power spectrum used in other sections. We add this here to provide a fair comparison of the constraints in the linear and NLO BAO cases. Overall we find it has a minor effect on the $M_\nu$ constraints and we can therefore justify neglecting it in the other calculations. Without IR resummation, the constraints in the NLO case in Figure \ref{fig_bao} improve by less than 10\% in the $\Lambda$CDM case and by an even lesser extent for all of the other cosmologies. For our \lq combined\rq~calculations, BAO make up only one component of the total constraining power so excluding IR resummation in that case should be reasonable. Finally, we note that, when only the BAO feature is used, the damping of the BAO signal can also be substantially mitigated through BAO reconstruction \cite{eisenstein/seo/etal:2007,2009PhRvD..79f3523P}, so the NLO constraints in Figure \ref{fig_bao} could be viewed as conservative. 

We follow \cite{ivanov_2018} to implement anisotropic BAO damping. When calculating the derivatives of the BAO component of the NLO galaxy power spectrum, we multiply the wiggles by a factor $\exp[-k^2\Sigma_{\rm tot}(\mu)^2]$ where

\begin{equation}
    \Sigma_{\rm tot}(\mu)^2 = (1+f\mu^2(2+f))\Sigma^2 + f^2\mu^2(\mu^2-1)\delta\Sigma^2
\end{equation}

\begin{equation}
    \Sigma^2 = \frac{1}{(2\pi)^3}\frac{4\pi}{3}\int_0^{k_S} dq P_{nw}(q)\left[1-j_0\left(\frac{q}{k_{osc}}\right)+2j_2\left(\frac{q}{k_{osc}}\right)\right]
\end{equation}

\begin{equation}
    \delta\Sigma^2 = \frac{1}{(2\pi)^3}4\pi\int_0^{k_S} dq P_{nw}(q)j_2\left(\frac{q}{k_{osc}}\right)
\end{equation}
$P_{nw}(k)$ is the baryon-CDM power spectrum without BAO (\lq no-wiggle\rq), $k_{osc}$ is the BAO scale $1/(110~h~\textrm{Mpc}^{-1})$, $k_S$ is set to 0.2 $h~\textrm{Mpc}^{-1}$, and $j_n$ are spherical Bessel functions of order $n$. 

\section{Discussion}\label{sec_discussion}

\subsection{Selection Effects}\label{subsec_selection_effects}

\begin{figure}
    \centering
    \includegraphics[width=0.8\textwidth]{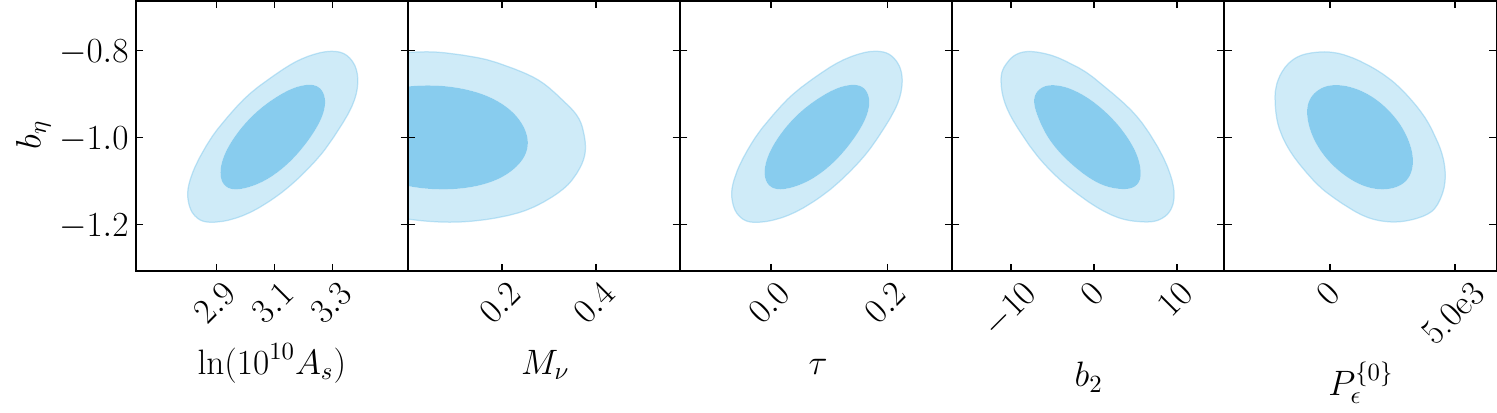}
    \caption{Contours showing the relationship between $b_\eta$ and a subset of the other parameters in our calculations, for the central redshift bin of the combined galaxy clustering Euclid forecast (without CMB lensing). Apart from $M_\nu$, these parameters include those with the most significant (anti-)correlation with $b_\eta$. Despite some degree of correlation between $b_\eta$, $A_s$ and $\tau$, the effect of varying $b_\eta$ on the constraint on $M_\nu$ is quite mild.}
    \label{fig:contour_b_eta}
\end{figure}

The observed galaxy power spectrum will always have a dependence on galaxy properties that alter the probability of a given galaxy being observed. Biases that arise in the observed power spectrum as a result of the specific subset of galaxies observed are called selection effects. Many selection effects are included by default in the bias expansion used here, but we have neglected selection effects that depend on the line-of-sight in this analysis. We will now justify this.

One particularly crucial selection effect arises from the dependence of the apparent brightness of a galaxy on its orientation with respect to us, as brighter galaxies are more likely to be detected. This selection effect is degenerate with the Kaiser effect \cite{hirata_tidal_2009,krause/hirata:2011}, and alters the value of $b_\eta$, which we fix to $-1$ when neglecting selection effects. An analogous effect arises when galaxies are selected on resonant line flux, such as Lyman-$\alpha$ \cite{zheng/etal:2011,wyithe/dijkstra:2011,Greig_2013}. 
We can briefly examine the impact of freeing $b_\eta$ to get an impression of its impact. 

Without any prior on $b_\eta$, in the combined case for $\Lambda$CDM, the NLO constraint on $M_\nu$ degrades from 0.044 eV to 0.051 eV. Imposing a $1\sigma$ prior on $b_\eta$ of 0.05 improves the constraint to 0.046 eV. Freeing $b_\eta$ does not have any significant effect on the BAO-only or free-streaming constraints. Figure \ref{fig:contour_b_eta} is a contour plot for $b_\eta$ and a small sample of other parameters in our calculations in our combined galaxy clustering forecasts (with marginalisation over the other parameters not shown). There is some clear degeneracy between $b_\eta$, $A_s$ and $\tau$, but the effect of $b_\eta$ on $M_\nu$ seems quite mild despite $M_\nu$ being correlated with the latter two. Of the nuisance parameters, $b_\eta$ shows the most clear degeneracy with $b_2$ and $P_\epsilon^{\{0\}}$.

We finally note that Euclid will also image the galaxies that they observe \cite{amendola_cosmology_2016}. This should allow for the intentional selection or reweighting of a galaxy sample in such a way as to reduce the line-of-sight dependent selection effects. 

\subsection{Varying the fiducial neutrino mass}\label{subsec_higher_mnu}

In Section \ref{subsec_free_streaming} we found that the scale-dependent suppression in the power spectrum caused by neutrinos is not particularly degenerate with any of the new parameters. The suppression of the power spectrum by neutrino free-streaming starts at larger scales than those at which the scale-dependent effects of the new bias parameters become significant. However, the free-streaming scale $k_{FS}$ increases with the neutrino mass. It is therefore worth investigating whether the constraints for a higher fiducial neutrino mass would be further weakened. 

We ran the combined results again for an increased $M_\nu$ of 0.27 eV assuming 3 degenerate mass states, and the constraints for all cosmologies were within a few percent of those obtained for the minimal neutrino mass. We conclude that any weakening of the constraints due to greater degeneracy with the non-linear bias parameters is compensated for by the increased amplitude of the suppression that comes with higher mass neutrinos. 

\subsection{Neutrino-Induced Bias}\label{subsec_neutrino_bias}

Something we have neglected in our treatment in this article is the effects of massive neutrinos themselves on the bias parameters. As we have seen, the free-streaming of massive neutrinos introduces a scale-dependence in the growth of perturbations, not just for neutrinos, but also cold dark matter and baryons. This should then result in some scale-dependence in the bias parameters. This was studied in detail by \cite{loverde_halo_2014} for $b_1$, who modified the spherical collapse and peak-background split derivations of the large-scale bias to account for this. Including this scale-dependent bias should work to counter the suppression caused by massive neutrinos, in principle weakening the signal. The same effect should also cause a scale-dependence in the higher-order bias parameters. 

This is a complicated topic and a full model of the next-to-leading-order power spectrum that accounts for this effect has yet to be developed. In addition, there is some disagreement on the precise size of the effect in N-body simulations \cite{castorina_cosmology_2013,castorina_demnuni_2016,villaescusa-navarro_imprint_2017,2018PhRvD..97l3526C,2019PhRvL.122d1302C}; see \cite{2019MNRAS.483..734R} for an estimate of the relevance of this effect for future surveys.
We therefore leave this for future work.

\subsection{Degeneracy between Bias Parameters}\label{subsec_bias_degeneracy}

We highlighted in Section \ref{subsec_free_streaming} that there is a significant anti-correlation between the two parameters associated with the tidal field, $b_{K^2}$ and $b_{td}$. An examination of the contributions of these terms to the power spectrum reveals why. 

Both $b_{td}$ and $b_{K^2}$ contribute to the $P_{gg}^{1-3}$ term. In \cite{desjacques_galaxy_2019}, the authors show that the contributions of the associated operators to $P_{gg}^{1-3}$ are linearly proportional. Their contributions are identical apart from a constant factor of 5/2, making the effects of varying the two bias parameters fully degenerate.

The $b_{K^2}$ parameter also contributes to the $P_{gg}^{2-2}$ term, but $b_{td}$ does not, preventing the two parameters from being perfectly anti-correlated. However, on inspection, we see that the relative change in $P_{gg}^{2-2}$ when varying $b_{K^2}$ is significantly smaller than that in $P_{gg}^{1-3}$, so changing $P_{gg}^{1-3}$ is the dominant effect of $b_{K^2}$. The parameters $b_{K^2}$ and $b_{td}$ therefore remain quite strongly anti-correlated, even after marginalising over other parameters. 

\subsection{Comparisons with Previous Work}\label{subsec_previous_work}

A recent article \cite{chudaykin_measuring_2019} performed an MCMC forecast for Euclid to 1-loop order in the power spectrum, as well as the tree-level bispectrum. Combining the current Planck likelihood with their Euclid forecast, they quote a constraint of 0.017 eV (excluding the bispectrum). Unlike us, they do not vary $N_{\scriptsize{\textrm{eff}}}$. When we fix $N_{\scriptsize{\textrm{eff}}}$, we obtain a constraint of 0.034 eV. There are two obvious differences between \cite{chudaykin_measuring_2019} and our work: 1) they use a much higher scale cut of $k_{\rm max} = 0.5 h\,{\rm Mpc}^{-1}$ but include a theoretical error which is a smooth increasing function of $k$; and 2) they use only six instead of our nine nuisance parameters. In their appendices, the authors show what happens if they impose a sharp cut-off $k_{\rm max}$ of 0.15 $h\,{\rm Mpc}^{-1}$, and their constraints degrade significantly, although they do not include CMB information in this case and do include the bispectrum, which we do not consider here, so it is not possible to perform a direct comparison. We also note that fixing $\tau$ in our own case takes the constraint from 0.034 eV to 0.023 eV, and a different constraint on $\tau$ may also contribute to the difference in results.

We note that in our work using the linear power spectrum only, our results using Fisher forecasts with sharp cut-offs in $k$ produced results that were remarkably consistent with MCMC forecasts with the controlled theoretical error approach (see the comparison to \cite{brinckmann_promising_2019} in \cite{boyle_understanding_2019}). 

The authors of \cite{chudaykin_measuring_2019} also examined the specific contribution of BAO to their final constraints, and found it to be modest. This emphasises our point in our previous work and here that neutrino mass constraints from BAO alone waste a great deal of usable information. 

\subsection{Biases on $M_\nu$ from the Power Spectrum Model}\label{subsec_fisher_bias}

\begin{figure}
\centering
\includegraphics[width=0.8\textwidth]{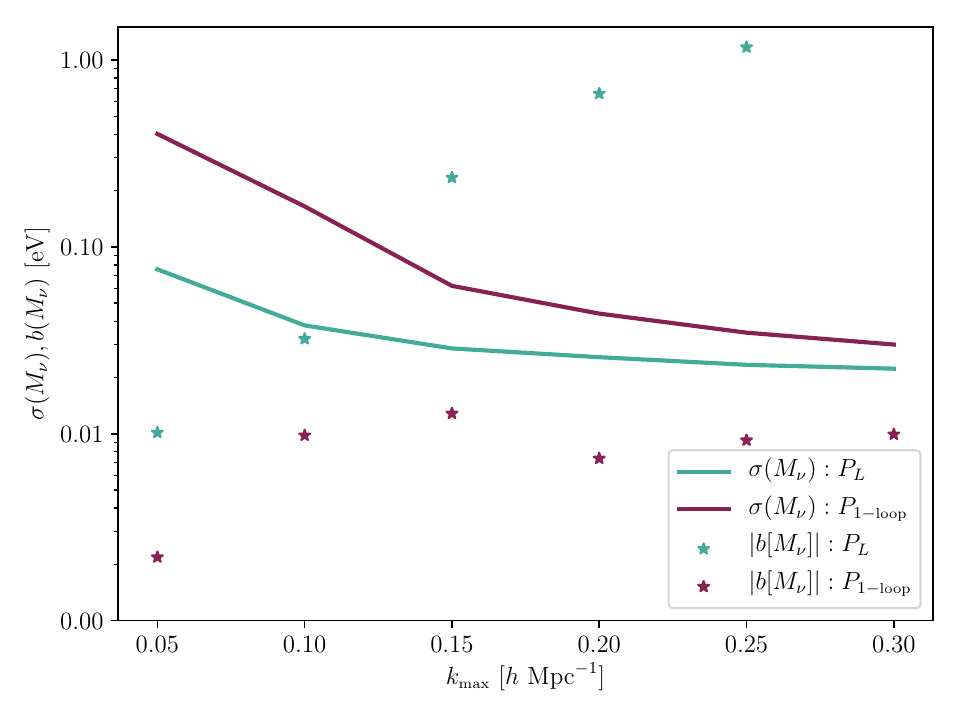}
\caption{The forecasted $1\sigma$ constraints on $M_\nu$ obtained using the linear (teal line) and 1-loop (purple line) galaxy power spectra for various $k_{\textrm{max}}$. The corresponding parameter biases resulting from not going to the next highest order in both cases are shown as points. The biases on $M_\nu$ in the linear case are actually negative. In the 1-loop case, the biases are relatively small and fluctuate between positive and negative depending on the cut-off scale. This explains why the bias does not monotonically increase with $k_{\textrm{max}}$. At $k_{\textrm{max}}=0.2~h
~\textrm{Mpc}^{-1}$, the value we use in our calculations, the 1-loop bias is approximately 20\% of the constraint.}\label{fig_fisher_bias}
\end{figure}

Using an inappropriate model for an observable (for example, using the linear power spectrum) will result in systematic biases in the parameters eventually constrained. Assume an observable $O(x)$ with observed values $\obs{O}$, where $x$ denotes the elements of the data vector, such as wavenumber and redshift bins, and cosmological parameters $\theta$ with true values $\bar{\theta}_i$. It is straightforward to extend the Fisher matrix formalism to estimate the magnitude of the biases produced by an incomplete model: 

\begin{equation}\label{eq_fisher_bias_result}
b[\theta_i] = \langle\thetasysi\rangle-\langle\bar{\theta}_i\rangle = \sum_j (F^{-1})_{ij} \sum_x \frac{\sys{O}(x)}{\textrm{Var}[\obs{O}(x)]}\pdv{\theory{O}(x)}{\theta_j},
\end{equation}

\noindent where $\thetasysi$ are the incorrectly inferred values of the cosmological parameters, $F$ is the Fisher matrix obtained using the incomplete model and $\partial \theory{O}/\partial \theta_j$ the corresponding derivatives. $\sys{O}$ is the systematic error in the model, i.e. the difference between the accurate and systematically incorrect fiducial models. 

For our example, we combine power spectra from multiple redshift bins to obtain the final result. Each redshift bin can be seen as an independent experiment. We also want to account for a CMB prior $F_{{\rm prior},ij}$ that is not affected by this systematic error. The bias for a given parameter as a result of using the linear galaxy power spectrum as opposed to the NLO power spectrum can then be calculated as

\begin{equation}\label{eq_fisher_bias_full}
b[\theta_i] = \langle\thetasysi\rangle-\langle\bar{\theta}_i\rangle = \sum_j \left(F_{\textrm{prior}}+\sum_z F_{z}\right)^{-1}_{ij}\sum_z \sum_k \frac{\sys{P_{gg}}(z,k)}{\textrm{Var}[P_{gg}(z,k)]}\pdv{P_{gg}(z,k)}{\theta_j}.
\end{equation}

\noindent Here, $\sys{P_{gg}}(z,k) = P_{gg,\scriptsize{\textrm{1-loop}}}(z,k)-P_{gg,L}(z,k)$. Figure \ref{fig_fisher_bias} shows the biases in the constraints on $M_\nu$ inferred from using the linear power spectrum, as a function of $k_{\textrm{max}}$. It is clear that the bias (shown as teal stars) significantly exceeds the linear power spectrum 1-$\sigma$ constraint (shown as a teal line) even at $k_{\textrm{max}}=0.1~h$/Mpc. These biases are also negative, as would be expected: the 1-loop power spectrum has enhanced clustering on small scales. Using the linear power spectrum to infer the neutrino mass would therefore make the power spectrum seem less suppressed on small scales, resulting in an inferred value of $M_\nu$ that is too small, resulting in the bias in Equation \ref{eq_fisher_bias_full} being negative.

Very conveniently, we require only the fiducial corrected power spectrum to estimate the bias from using the incorrect model in Equation \ref{eq_fisher_bias_full}. This means we can also make an estimate of how much our results would be affected by not moving to the next order, 2-loop power spectrum. We make use of the 2-loop matter power spectrum calculated for \cite{agarwal_information_2020} and model the 2-loop contribution to the galaxy power spectrum as approximately $P_{gg,\textrm{2-loop}}=(b_1+f\mu^2)^2P_{m,\textrm{2-loop}}(k)$. It is clear that the parameter biases are much less significant and the 1-loop power spectrum is much more robust up to higher $k_{\textrm{max}}$. The bias is about 20\% of the error at $k_{\textrm{max}}=0.2~h~\textrm{Mpc}^{-1}$.

The above result for the bias on the 1-loop power spectrum is intended to be approximate. The two-loop matter power spectrum used is for a slightly different cosmology and is only calculated at a single redshift and rescaled (so does not include the redshift-dependent effects of neutrinos). More importantly, it also neglects higher-order bias terms arising in the two-loop galaxy power spectrum, whose coefficients could be relatively less suppressed at higher redshifts.

\subsection{Alternative Galaxy Surveys}\label{subsec_alternative_galaxy_surveys}

One might ask how the conclusions derived for the Euclid-like survey in this work could change with the survey parameters. The Roman Telescope (formerly WFIRST) survey \cite{green_wide-field_2012} provides a natural complement to the Euclid survey for examination, having a considerably smaller area (2000 $\deg^2$ in our model) but probing deeper redshifts (up to $z=2.8$). 

Our model for the Roman telescope is given in Table \ref{table_wfirst} of Appendix \ref{app_surveys}. We have updated our model to match that provided as Design Reference Mission (DRM) 1 in \cite{green_wide-field_2012}, so the results should not be directly compared with those for \lq WFIRST\rq~provided in \cite{boyle_deconstructing_2018, boyle_understanding_2019}, particularly as the area of the survey has been updated from 2000 to 3400 $\deg^2$. The total number of galaxies here is approximately 16.5 million but with a greater redshift depth than Euclid (up to $z=2.7$). Note that the survey model is also different to that analysed in \cite{agarwal_information_2020}, which focused exclusively on the H$\alpha$-emitting galaxies. 

At $k_{\textrm{max}}=0.2~h~\textrm{Mpc}^{-1}$, while Euclid gave $\sigma(M_\nu)$ = 0.042 eV with a bias of 0.008 eV relative to the 2-loop model, the Roman Telescope gives $\sigma(M_\nu)$ = 0.078 eV with a bias of 0.0028 eV. The approximate factor of two decrease in the constraint is accompanied by an improvement in the bias, which is probably a result of the increased redshift range, as the relative importance of nonlinear corrections to the power spectrum shrinks with increasing $z$. Increasing $k_{\textrm{max}}$ to $0.35~h~\textrm{Mpc}^{-1}$ is required to reduce the error on $M_\nu$ to that of Euclid, achieving a forecasted error for the Roman Telescope of 0.042 eV with a bias of 0.0032. 

\section{Conclusions}\label{sec_conclusions}

The aim of this paper has been to provide an indication of how forecasted neutrino mass constraints from galaxy clustering and CMB surveys are altered on moving to next-to-leading-order power spectra. The results are relatively encouraging considering that the next-to-leading-order galaxy power spectrum implemented here introduces 7 new unconstrained nuisance parameters. When performing Fisher forecasts for galaxy clustering for a Euclid-like survey and CMB lensing from an experiment like Simons Observatory, we find:

\begin{itemize}
\item When the full redshift-space galaxy power spectrum is used, the constraints on $M_\nu$ in the $\Lambda$CDM case degrade from 0.026 eV in the linear case to 0.044 eV in the NLO case. This degrades to 0.12 eV if both curvature and the dark energy equation of state are also allowed to vary. The degeneracy between $M_\nu$ and $\tau$ becomes much less significant in the NLO case, as the error on $M_\nu$ is primarily inflated by uncertainties on the new nuisance parameters. Adding CMB lensing information tightens the NLO constraint to 0.034 eV by helping reduce the level of degeneracy with the nuisance parameters, and the connection between the errors on $\tau$ and $M_\nu$ then becomes significant once again.
\item When only distance information from BAO are used and combined with CMB lensing, the constraints are approximately the same in the linear and NLO cases: 0.048 eV in the $\Lambda$CDM case and 0.82 eV when curvature and the dark energy equation of state are left free.
\item When only measurements of the direct free-streaming effect in the power spectra are used (the scale-dependence in the matter power spectrum and the structure growth rate), the constraints on $M_\nu$ are independent of $\Omega_k$ and $w$ and approximately 0.11 eV in all cases (to be compared to 0.08 eV for the linear power spectra). 
\end{itemize}

CMB lensing plays a much more important role here than in \cite{boyle_understanding_2019}, where we examined the linear case, but the effect of adding the cross-correlation between galaxy clustering and CMB lensing remains largely negligible in the constraints. 

In Section \ref{subsec_fisher_bias}, we attempted to quantify the biases introduced in predictions of the neutrino mass by assuming an incomplete model for the power spectrum. We showed that the bias resulting from assuming the linear galaxy power spectrum already almost exceeds the predicted constraint on $M_\nu$ at at $k_{\rm max}$ of 0.1 $h~\textrm{Mpc}^{-1}$. An error estimate for the 1-loop power spectrum, furnished by the 2-loop matter power spectrum, shows that the former remains robust to much higher $k_{\rm max}$, with a bias of approximately 20\% of the $M_\nu$ constraint at $k_{\rm max}$ of 0.2 $h~\textrm{Mpc}^{-1}$.

In terms of future work, we note that combined analysis with the bispectrum, which we do not consider here, could help break degeneracies \cite{Hahn_2020} and better constrain the bias parameters. The effects of massive neutrinos themselves on the bias parameters should also be analysed (see Section \ref{subsec_neutrino_bias}). 

Finally, we emphasise that constraints on the mass of neutrinos from cosmology are generally highly sensitive to the assumed cosmological model, and should be quoted with the appropriate caveats. 

\acknowledgments

We would like to thank Eiichiro Komatsu for providing feedback on the first draft of this paper, which appeared as a chapter in AB's doctoral thesis. We would also like to thank Nishant Agrawal, for helping so generously with code and results comparisons. Finally we would like to thank Donghui Jeong and Vincent Desjacques for useful discussions and assistance. FS acknowledges support from the Starting Grant (ERC-2015-STG 678652) ``GrInflaGal'' of the European Research Council.

\bibliographystyle{JHEP}

\bibliography{neutrino_3}

\appendix

\newpage

\section{Survey Parameters}\label{app_surveys}

Tables \ref{table_euclid} and \ref{table_wfirst} give the survey parameters for Euclid and the Roman Telescope, respectively. The final columns give the estimated halo Lagrangian radii used to determine the fiducial values of $\blapl$ and $\beta_{\nabla^2v}$.

\begin{table}[h]
    \centering
    \begin{tabular}{c c c c c}
    \hline
$z$ & $V$ ($h^{-3}$ Gpc$^{3}$) & $10^3\times n_g$ (Mpc$^3$ $h^{-3}$) & $b_1$ & $R_L$ (Mpc $h^{-1}$)\\
\hline
0.65 & 2.59 & 0.637 & 1.07 & 2.41 \\
0.75 & 3.07 & 1.441 & 1.12 & 1.88 \\
0.85 & 3.52 & 1.619 & 1.17 & 1.79 \\
0.95 & 3.93 & 1.489 & 1.23 & 1.8 \\
1.05 & 4.29 & 1.32 & 1.28 & 1.83 \\
1.15 & 4.62 & 1.144 & 1.34 & 1.84 \\
1.25 & 4.9 & 0.995 & 1.39 & 1.86 \\
1.35 & 5.14 & 0.832 & 1.45 & 1.9 \\
1.45 & 5.35 & 0.659 & 1.51 & 1.94 \\
1.55 & 5.52 & 0.503 & 1.57 & 2.0 \\
1.65 & 5.66 & 0.364 & 1.62 & 2.08 \\
1.75 & 5.78 & 0.253 & 1.68 & 2.18 \\
1.85 & 5.88 & 0.166 & 1.74 & 2.28 \\
1.95 & 5.95 & 0.101 & 1.8 & 2.43 \\
2.05 & 6.01 & 0.037 & 1.86 & 2.72 \\

\hline
    \end{tabular}
    \caption{Survey parameters assumed for Euclid.}
    \label{table_euclid}
\end{table}

\begin{table}[h]
    \centering
    \begin{tabular}{c c c c c}
    \hline
$z$ & $V$ ($h^{-3}$ Gpc$^{3}$) & $10^3\times n_g$ (Mpc$^3$ $h^{-3}$) & $b_1$ & $R_L$ (Mpc $h^{-1}$)\\
\hline

1.35 & 1.08 & 1.253 & 1.44 & 1.72 \\
1.45 & 1.13 & 1.24 & 1.48 & 1.68 \\
1.55 & 1.17 & 1.18 & 1.52 & 1.65 \\
1.65 & 1.21 & 1.091 & 1.56 & 1.65 \\
1.75 & 1.24 & 1.007 & 1.6 & 1.63 \\
1.85 & 1.26 & 0.944 & 1.64 & 1.62 \\
1.95 & 1.28 & 0.961 & 1.68 & 1.58 \\
2.05 & 1.3 & 0.988 & 1.72 & 1.52 \\
2.15 & 1.31 & 1.021 & 1.76 & 1.47 \\
2.25 & 1.32 & 1.001 & 1.8 & 1.44 \\
2.35 & 1.33 & 0.858 & 1.84 & 1.44 \\
2.45 & 1.33 & 0.718 & 1.88 & 1.46 \\
2.55 & 1.33 & 0.552 & 1.92 & 1.5 \\
2.65 & 1.33 & 0.416 & 1.96 & 1.53 \\

\hline
    \end{tabular}
    \caption{Survey parameters assumed for the Roman Telescope.}
    \label{table_wfirst}
\end{table}

\end{document}